\documentclass[letterpaper, 10 pt, conference]{ieeeconf}

\IEEEoverridecommandlockouts                              

\usepackage{cite}
\usepackage{amsmath}
\usepackage{amssymb}
\usepackage{amsfonts}
\usepackage{graphicx}
\usepackage{textcomp}
\usepackage{xcolor}

\usepackage{cancel}
\usepackage{subcaption}

\usepackage{amsthm}

\usepackage{siunitx}
\usepackage{cleveref}

\newcommand{\norm}[1]{\left|#1\right|}
\newcommand{\flow}{\mathcal{F}}
\newcommand{\jump}{\mathcal{J}}
\newcommand{\real}{\mathbb{R}}
\newcommand{\sign}[1]{\text{sign}\left(#1\right)}
\newcommand{\Max}[1]{\max\left(#1\right)}	
\newcommand{\diag}[1]{\text{diag}\left(#1\right)}

\newcommand{\eref}[1]{(\ref{#1})}
\newcommand{\fref}[1]{Fig. \ref{#1}}

\theoremstyle{plain}
\newtheorem{theorem}{Theorem}

\theoremstyle{plain}
\newtheorem{assumption}{Assumption}
\newtheorem{corollary}{Corollary}[assumption]

\theoremstyle{remark}
\newtheorem{remark}{Remark}

\title{\LARGE \bf
	Adaptive Feedforward Control For Reset Feedback Control Systems -\\ \large{Application in Precision Motion Control}
}

\author{Karst Brummelhuis$^1$, Niranjan Saikumar$^2$, Jan-Willem van Wingerden$^1$, S. Hassan HosseinNia$^{2*}$ 
	\thanks{$^1$ Delft Center for Systems and Control, 3ME, TU Delft}
	\thanks{$^2$ Precision and Microsystems Engineering, 3ME, TU Delft, the Netherlands}
	\thanks{$^{*}$Corresponding author - {S.H.HosseinNiaKani@tudelft.nl}}
}
    
\begin{document}
	
\maketitle

\begin{abstract}
This paper presents a novel adaptive feedforward controller design for reset control systems. The combination of feedforward and reset feedback control promises high performance as the feedforward guarantees reference tracking, while the non-linear feedback element rejects disturbances. To overcome inevitable model mismatches, the feedforward controller adapts to increase precision in reference tracking. Where linear existing adaptive feedforward controllers do not guarantee convergence in the presence of reset, this work presents a novel adaptive law based on converging and diverging regions of adaptation to achieve good tracking. Experimental results demonstrate the claimed advantage of the novel method.
\end{abstract}

\begin{keywords}
adpative feedforward, reset control, mechatronics, nonlinear control, motion control
\end{keywords}

\section{Introduction}
The precision industry with applications such as atomic force microscopes, photo-lithography machines etc. continues to push the limits in terms of precision and throughput. As the control precision and bandwidth contribute to the error margins of production and throughput, improved control strategies have a direct impact on increasing profit and production quality. As an industry-standard, PID control is still preferred in most applications, given the fact that more than 90\% of all control loops are PID \cite{samad2019ifac}. However, PID is fundamentally limited by its linearity. Limitations which can be expressed by the waterbed effect can be suppressed by introducing a resetting integrator. The idea of reset was introduced in 1958 by J.C. Clegg \cite{Clegg}. Resetting the integrator whenever a zero crossing of the input occurs, yields a significant reduction in phase lag. This means that introducing non-linear reset control favours the freedom of design and improves the stability margins. Recent research on reset control can be found in \cite{GSORE,MagneticFORE,ResetDistObserver,FractionalReset,ComplexOrder,ResetTransfer}.

Since describing function analysis allows reset control to be used in the loop shaping framework, the usage of reset control is highly desired compared to other non-linear control methods such as sliding-mode control, backstepping control, or model predictive control. However, reset control causes higher-order harmonics to be produced in the output of the controller. In the system response, these harmonics play a role by causing unwanted behaviour, which results in poor performance in terms of tracking. To achieve better tracking, control strategies are required to handle the higher-order harmonics. Strategies like PI+CI\cite{PICI}, reset band\cite{resetband}, partial reset\cite{resetcontrolbook} partially solve this problem, but undesirably deduct part of favourable properties of pure reset control, and hence only provide a trade-off. 

Perfect reference tracking can be achieved with feedforward control\cite{mechatronics}. With exact knowledge of the plant parameters, the feedforward element can be tuned such that output exactly follows the reference. In reality, exact knowledge of the plant is practically impossible due to non-linearities and uncertainties. As feedforward accounts for tracking, the main purpose of the feedback controller is disturbance rejection and compensation of plant uncertainties. Since reset control yields higher phase margin compared to linear control, a combination of feedforward with reset control promises high performance in both stability and tracking.
However, as fixed feedforward is sensitive to model mismatches, a dynamic feedforward model is desired. Adaptive or iterative feedforward update laws are suggested to on-line adjust the feedforward to reduce the error induced by model mismatches. Non-linearities introduced by the reset feedback element exclude the direct implementation of linear adaptive or iterative laws \cite{FEL},\cite{ILC}, \cite{datadriven}.

Combinations of adaptive feedforward with reset control are explored in literature.  Adaptive feedforward with first-order reset elements in the control loop is researched in \cite{EGR} and \cite{powersplittransmission}. The research which is done in \cite{EGR} only accounts for set-point regulation. While this limitation is solved in \cite{powersplittransmission}, the method is only designed for first-order plants controlled by first-order reset elements. However, for application in precision motion control, in which general systems are represented by a mass-spring-damper system or a cascade of mass-spring-damper systems, none of the existing methods is applicable.

The main contribution of this paper is the adaptive feedforward controller for general reset feedback control systems, applicable on plants of any order. The adaptive algorithm handles the non-linearities introduced by the reset element and adapts the feedforward parameters to achieve more accurate tracking. Different types of reset elements are accommodated within the framework. Also, by implementing the algorithm on  an experimental setup, it is demonstrated that the adaptive feedforward converges and high tracking precision is achieved.

\Cref{sec:feedforward} provides necessary information on feedforward in general, with \Cref{sec:reset} providing the preliminaries for reset control. In \Cref{sec:algorithm} the adaptive feedforward algorithm for general reset control systems is elaborated. The developed algorithm is implemented on a precision positioning stage and the results are presented in \Cref{sec:application}. Conclusions and recommendations for further work are given in \Cref{sec:conclusion}. 

\section{Control structure} \label{sec:feedforward}
\begin{figure}[htbp!]
	\centering
	\includegraphics[width=\linewidth]{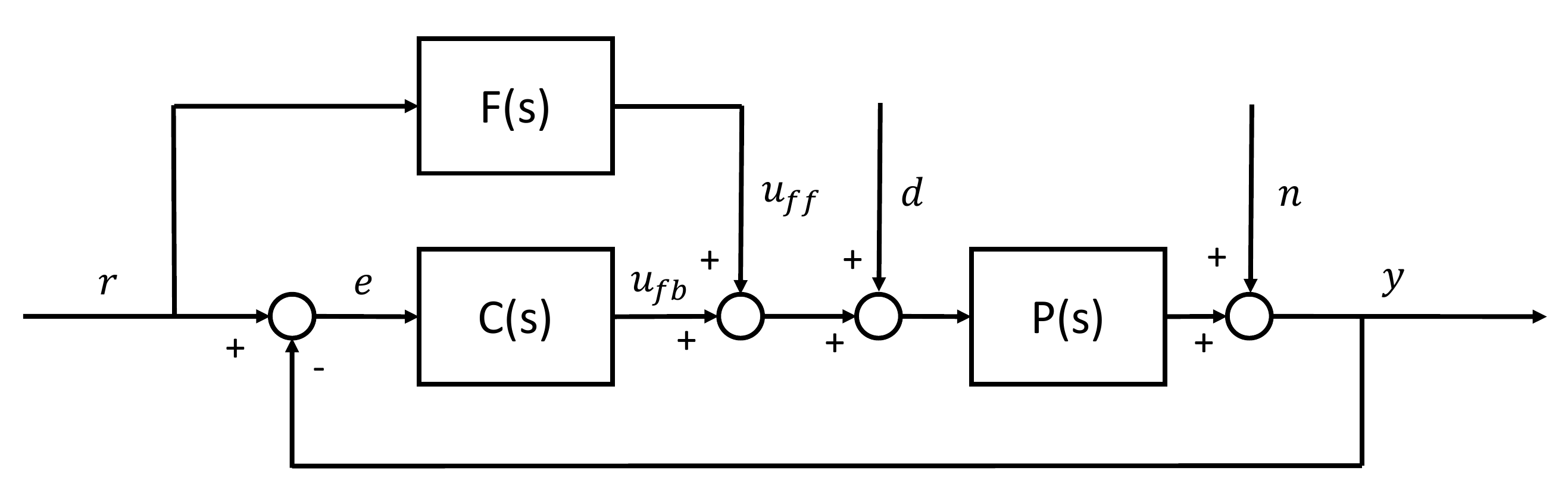}
	\caption{Two degree of freedom closed loop structure}
	\label{fig:blockdiagram}
\end{figure}

The general structure of a control scheme involving both feedforward and feedback control is given in the block diagram in \fref{fig:blockdiagram}. 
The dynamics of this system are governed by the transfer function given in \eref{eq:transfer}, 
\begin{equation}
	y = \dfrac{PC+PF}{PC+1}r + \dfrac{P}{PC+1}d + \dfrac{1}{PC+1}n,
	\label{eq:transfer}
\end{equation}
where the plant $P$ and feedback controller $C$ are both SISO. The feedforward controller dynamics are governed by $F$. The exogenous signals are the reference $r\in\real$, external disturbance $d\in\real$, and noise $n\in\real$. The feedforward controller only affects the reference tracking properties of the plant, hence disturbance rejection properties of the feedback controller are maintained for any arbitrary feedforward controller. Ideal feedforward is achieved by the inverse of the plant dynamics $F=P^{-1}$, as the mapping from the reference to the output becomes unity.
\begin{equation}
	y = \dfrac{PC+PP^{-1}}{PC+1}r 
\end{equation}
The accuracy of a fixed feedforward is directly influenced by the quality of the model used for plant inversion. Accuracy of the physical model is influenced by environmental disturbances, non-linearities of the plant, model uncertainties, or time-varying parameters such as added mass. Adaptive or iterative feedforward models can estimate correct model parameters over time. 

For reset control systems, existing linear adaptive or iterative feedforward controllers are not directly applicable for two reasons:
\begin{enumerate}
	\item As reset control consists of a flow and a jump state, only the flow is considered in existing linear laws.
	\item Reset control can be implemented with an unstable base linear system.
\end{enumerate}
This means that linear laws based on a stable linear closed-loop, do not guarantee convergence. Even in the case where the base linear system is stable, large spikes in control value can be observed at reset instants. Linear methods do not account for this behaviour, and therefore are unlikely to solve the adaptation problem.

\section{Reset Control} \label{sec:reset}
\subsection{General Reset Control}
In this section, reset control is introduced and several commonly used reset elements are presented. These different types of reset elements can be implemented as part of a feedback controller in the framework presented in \fref{fig:blockdiagram}. General reset element dynamics are governed by: 
\begin{equation}
	\begin{aligned}
		\dot{x}_r(t) &= A_rx_r(t) + B_re(t) & \left(x_r,e\right)&\in\flow \\
		x_r(t^+) &= A_\rho x_r(t) & \left(x_r,e\right)&\in\jump \\
		u(t) &= C_rx_r(t) + D_re(t)
	\end{aligned}
	\label{eq:statereset}
\end{equation}
where the matrices $A_r$, $B_r$, $C_r$, and $D_r$ describe the state space of the reset element, $e(t)\in\real$ is the error input, $x_r(t)\in\real^{n_r}$ are the states, and $u(t)\in\real$ is the controller output. The linear dynamics given by first and third equations of \eqref{eq:statereset} are referred to as the base linear system. The controller states propagate according to the base linear system if $\left(x_r,e\right)\in\flow$. Whenever the reset conditions are met, $\left(x_r,e\right)\in\jump$, specified controller states are reset according to the reset matrix $A_\rho$. In this context $\flow$ and $\jump$ represent the flow set and jump set, respectively. Several sets of reset conditions are presented in literature. Dominant reset conditions in literature are either based on zero crossings of the error\cite{Hbeta}, or based on the signs of the reset controller states and plant output\cite{stabilityzaccarian}. This work focusses on reset elements based on zero crossings of the error. So,
\begin{equation}
	\begin{aligned}
		\flow &:= \left\lbrace \left(x_r,e\right)\in\real:e\neq0 \right\rbrace\\ 
		\jump &:= \left\lbrace \left(x_r,e\right)\in\real:e=0 \right\rbrace
	\end{aligned}
\end{equation}

\subsection{Reset elements}
Different types of reset elements have been presented in literature. There are three main configurations for reset elements, the Clegg Integrator, first-order reset element, and second-order reset element. 
\subsubsection{Clegg Integrator}
The Clegg integrator (CI) was first introduced as a resetting technique in \cite{Clegg}. Using a Clegg integrator yields the advantage of 52\textdegree \ in reduced phase lag compared to a linear integrator. For the CI we have the following system matrices for \eref{eq:statereset}:
\begin{equation*}
	A_r = 0, \quad B_r = 1 \quad C_r = 1 \quad D_r = 0 \quad A_\rho = 0
\end{equation*}
In literature, a reset element is expressed in transfer function form using an arrow crossing through the function.
\subsubsection{First Order Reset Element}
The first extension of the CI is the First Order Reset Element (FORE)\cite{FORE1975} which behaves as the reset equivalent of a first order LPF with corner frequency $\omega_r$. The system matrices for FORE are given by:
\begin{equation*}
	A_r = -\omega_r \quad B_r = \omega_r \quad C_r = 1 \quad D_r = 0 \quad A_\rho = 0
\end{equation*}
\subsubsection{Second Order Reset Element}
A further extension of the CI and FORE is the Second Order Reset Element (SORE) \cite{SORE}. By increasing the order, the ability arises of designing resetting notch filters or second-order low-pass filters with the extra design variable, damping factor $\beta_r$. The system matrices for SORE are given by:
\begin{equation*}
	\begin{aligned}
		&A_r = \begin{bmatrix}
			0 & 1 \\ -\omega_r^2 & -2\beta_r\omega_r
		\end{bmatrix}, \quad B_r = \begin{bmatrix}
			0 \\ \omega_r^2
		\end{bmatrix}, \\ &C_r = \begin{bmatrix}
			1 & 0
		\end{bmatrix}, \quad D_r = 0, \quad A_\rho = 0
	\end{aligned}
\end{equation*}

Adapted versions of the reset elements are researched to enhance the capabilities of reset control. Examples of these adapted versions are generalized FORE and SORE, and also the CgLp element, all mentioned below. 

\subsubsection{Generalized FORE and SORE \cite{frequencydomain,GSORE}}
Instead of resetting to zero, the generalized reset element resets the controller states to a fraction of their pre-reset value according to the value of $\gamma$. The reset matrices, $A_\rho$, for the GFORE and GSORE are respectively given by: 
\begin{equation}
	A_\rho = \gamma \qquad A_\rho = \gamma I_{2\times 2}
\end{equation}
For open-loop stability and convergence, $A_\rho$ needs to be Schur stable\cite{frequencydomain}.

\subsubsection{Constant in Gain Lead in Phase}
The Constant in gain-lead in phase element(CgLp) is introduced in \cite{GSORE}. This element is used to provide phase lead over a broadband frequency range, whereas other reset elements only provide a reduction of phase lag. CgLp makes use of a GFORE/GSORE in combination with a corresponding order linear lead filter. In transfer function domain the GFORE based CgLp is given by the combination of reset filter and linear filter as given in \eref{eq:CgLp1tf}, with $\omega_{r\alpha}<\omega_r<<\omega_f$. $\omega_{r\alpha}$ accounts for the shift in corner frequency introduced by the resetting action and the magnitude of shift determined by $\gamma$.
\begin{equation}	
	C(s) = \dfrac{1}{\cancelto{}{s/\omega_{r\alpha} + 1}} 
	\qquad \dfrac{s/\omega_r + 1}{s/\omega_f + 1}
	\label{eq:CgLp1tf}
\end{equation}

\subsection{Closed loop and stability}
For Lyapunov stability analysis, the closed-loop dynamics of the system including the feedback reset controller are required. The closed-loop dynamics of a system controlled by a reset element are given in \eref{eq:closedloop}. For ease of notation $x(t)$ is represented as $x$. The plant states and controller states are indicated by $x_p\in\real^{n_p}$ and $x_r\in\real^{}$, respectively. The considered plant is a SISO system.
\begin{equation}
	\begin{aligned}
		\begin{bmatrix}
			\dot{x}_p \\ \dot{x}_r
		\end{bmatrix} &= \begin{bmatrix}
			A_p - B_pD_rC_p & B_pC_r \\ -B_rC_p & A_r
		\end{bmatrix}\begin{bmatrix}
			x_p \\ x_r
		\end{bmatrix} &+ \begin{bmatrix}
			B_pD_r \\ B_r
		\end{bmatrix}r \\
	& & \left(x,r\right) \in\flow \\
		\begin{bmatrix}
			x_p^+ \\ x_r^+
		\end{bmatrix} &= \begin{bmatrix}
			I & 0 \\ 0 & A_{\rho}
		\end{bmatrix}\begin{bmatrix}
			x_p \\ x_r
		\end{bmatrix} & \left(x,r\right) \in\jump \\
		y &= \begin{bmatrix}
			C_p & 0
		\end{bmatrix}
	\end{aligned}
	\label{eq:closedloop}
\end{equation}
To avoid Zeno solutions, the following assumption is made.
\begin{assumption}
	For the closed-loop reset control system, the states after reset are such that immediate successive reset is not possible.
	\begin{equation}
		x\in\jump \Rightarrow x^+\in\flow
	\end{equation}
\end{assumption}
Note that for discrete systems, Zeno behaviour is excluded by definition, as resets can only occur at every sampling instant. 

A homogeneous Lyapunov function has been used in \cite{Hbeta} to obtain the $H_\beta$ condition for stable base linear systems. And for all such systems, the main condition which needs to be satisfied for proving asymptotic stability is given as the one below.

\begin{theorem}\cite{Hbeta}
	Let $V(x):\real^n\rightarrow\real$, where $n=n_p+n_r$, be a continuously differentiable, positive-definite, radially unbounded function, 
	such that:
	\begin{equation}
		\begin{aligned}
			\dot{V}(x) &<0 & x&\in\flow, x \neq 0 \\
			\Delta V(x) &= V(A_\rho x)-V(x)\leq0 & x&\in\jump
		\end{aligned}
		\label{eq:Lyap1}
	\end{equation}
	If both conditions hold, the equilibrium $x=0$ is globally asymptotically stable. 
	\label{th:Lyapunov1}
\end{theorem}

\section{Adaptive Feedforward Algorithm} \label{sec:algorithm}

In this section the novel adaptive feedforward controller is introduced. \textit{The adaptive algorithm is based on characteristic properties of reset feedback controllers.}

\subsection{Important properties of the reset control loop}
The time in the flow set $\flow$ after a reset is defined by $\tau$. At a reset instant, this time variable is reset to zero.
\begin{equation*}
	\dot{\tau}(t) = 1  \ \ \left(x,r\right)\in\flow, \qquad \tau(t^+) = 0  \ \ \left(x,r\right)\in\jump
	\label{eq:tau}
\end{equation*}
\begin{assumption}
	The closed-loop with the reset feedback controller of \eref{eq:closedloop}, excluding the feedforward input, is asymptotically stable. 
	\label{assump:cl}
\end{assumption}
\begin{corollary}
	Because of the asymptotic stability of the closed-loop, when $x\in\flow$ a reset will happen after a finite amount of time.
	\label{cor:flowreset}
\end{corollary}

\begin{remark}
	Consider the case with a reset feedback controller, where the feedforward signal is not present. A reset controller can either have a stable or unstable base linear system. If unstable, the resetting action can stabilize the closed-loop\cite{EGR},\cite{powersplittransmission}. This resetting action introduces additional stability, which can be understood as bringing the controller states closer to the required value when the control input is diverging from the ideal value in the flow interval after some time $\overline{\tau}$. Due to the sharp nature of reset, the feedback control input will only enter the converging region after some amount of time $\underline{\tau}$.  Hence it can be stated that in every flow interval, the control input to the plant is converging towards ideal control input $u^*$ in the interval $[\underline{\tau},\overline{\tau}]$. Note that for both a stable and unstable base linear system this interval exists.
\end{remark}

\begin{assumption}
	In every flow set there exists a region with a lower bound $\underline{\tau}$ and an upper bound $\overline{\tau}$ in which $u_{fb}$ converges to $u^*$:
	\begin{equation*}
		u_{fb}(\tau)\rightarrow u^*(\tau) \qquad \forall \  \underline{\tau}<\tau<\overline{\tau}
	\end{equation*}
	\label{assump:ufb}
\end{assumption}
\vspace{-20pt}
The overall control system consists of a feedforward and a feedback controller. 
\begin{corollary}
	In the case of both feedforward and feedback, $u_{fb}+u_{ff}$ converges to $u^*$ within the region lower bounded by $\underline{\tau}$ and upper bounded by $\overline{\tau}$:
	\begin{equation*}
		u_{fb}(\tau)+u_{ff}(\tau)\rightarrow u^*(\tau) \qquad \forall \  \underline{\tau}<\tau<\overline{\tau}
	\end{equation*}
	\label{cor:utotal}
\end{corollary}
\vspace{-30pt}

\subsection{Parametrization}
The adaptive feedforward controller is designed for the poles of the plant. The zeros are not considered and therefore not estimated. However, when the plant has zeros, this can be considered as plant uncertainties which are to be handled by the feedback reset controller. For plants without zeros, the inverse is given by: 
\begin{equation}
	\tilde{P}^{-1}(s) = \theta_n s^{n-1} + \theta_{n-1} s^{n-2} + \hdots + \theta_1
	\label{eq:pinv}
\end{equation}
where, as a result, the feedforward control input is defined as $\tilde{P}^{-1}(s)r$. To define an update law, the feedforward signal is parametrised in linear form. The parametrization consists of a vector $\theta\in\real^{n_p}$ containing the individual gains, referred to as the feedforward parameters and a vector $\phi\in\real^{n_p}$ containing reference information. The size of $\theta$ is determined by the order of the plant.
\begin{equation}
	\theta = \begin{bmatrix}
		\theta_n & \theta_{n-1} & \hdots & \theta_1
	\end{bmatrix}^T
\end{equation}
Containing the same number of elements, $\phi$ consists of the reference and its first $n$ derivatives:
\begin{equation}
	\phi = \begin{bmatrix}
		r^{(n)}& r^{(n-1)} & \hdots & r
	\end{bmatrix}^T
	\label{eq:phia}
\end{equation}
In most high precision motion systems, the reference is designed by taking the velocity, acceleration, jerk and even snap limitations of the system, ensuring that these derivatives are available \cite{lambrechts2005trajectory}.
With these two vectors, the feedforward signal is constructed in real-time as:
\begin{equation}
	u_{ff} = \theta^T\phi
\end{equation}
In the case where the derivative information of  the reference is not completely known beforehand, \eref{eq:phib} can be used, where $\Lambda(s)$ is a stable low-pass filter of order $n$:
\begin{equation}
	\phi = \begin{bmatrix}
		\dfrac{s^n}{\Lambda(s)} & \dfrac{s}{\Lambda(s)} & \hdots & \dfrac{1}{\Lambda(s)}
	\end{bmatrix}^T r \label{eq:phib}
\end{equation}
The update algorithm defined in the following section is based on this parametrization.

\subsection{Update algorithm}
The error on the feedforward control input is defined by:
\begin{equation}
	\epsilon(t) = \theta^{*T}\phi(t)-\theta^{T}\phi(t)
	\label{eq:epsilon}
\end{equation}
where $\theta^{*T}\phi=u^*$ is the ideal control input and $\theta^*$ is the vector of ideal feedforward parameters. Based on \Cref{cor:utotal}, $\theta^{*T}\phi$ is represented by the total control input. With the definition of
\begin{equation}
	\epsilon(\tau) = u_{fb}(\tau) \qquad \forall \  \underline{\tau}<\tau<\overline{\tau}
	\label{eq:epsilon}
\end{equation}
a gradient descent based update law for $\theta$ is defined. This update algorithm is given as follows:
\newlength{\temp}
\settowidth{\temp}{$\theta(t^+) = \theta(t) + \Gamma_J\norm{J(t)}\theta_J(t)$}
\begin{equation}\label{eq:updatelaw}
	\begin{aligned}
		\begin{aligned}
			&\makebox[\temp][l]{$\dot{\theta}(t) = \Gamma_F\psi(t) u_{fb}(t)\phi_n(t)$} \\
			&\dot{\theta}_J(t) = \psi(t) u_{fb}(t)\phi_n(t) \end{aligned}
		&& \left(x,r\right)&\in\flow \\
		\begin{aligned}
			&\theta(t^+) = \theta(t) + \Gamma_J\norm{J(t)}\theta_J(t)\\
			&\theta_J(t^+) = 0 \end{aligned}
		&& \left(x,r\right)&\in\jump
	\end{aligned}
\end{equation}
where $\Gamma_{F_{n_p\times n_p}}$ and $\Gamma_{J_{n_p\times n_p}}$ are the adaptive gain matrices, with positive diagonal terms. These diagonal matrices can be tuned to determine the rate of adaptation of the feedforward parameters. $J(t) = u_{fb}(t^+) -u_{fb}(t)$ defines the jump in control value at reset. $\phi_n(t)$ is the normalized vector of the filtered input signals contained in $\phi(t)$. Normalization ensures the boundedness of the signal. The normalization matrix $P\geq0$ and $\zeta>0$ can be set according to the reference properties.
\begin{equation}
	\phi_n(t) = \dfrac{\phi(t)}{\phi(t)^TP\phi(t)+\zeta}
\end{equation}
Let us define $\psi(t)\in\lbrace 0,1\rbrace$ as a variable which determines when the controller is in the converging region: 
\begin{equation}
	\psi(t) = \Max{\beta,\xi(t)}
	\label{eq:psi}
\end{equation}
Here $\beta$ is a design variable and $\xi(t)$ is the filtered absolute tracking error. With $\beta \in \lbrace 0,1\rbrace$ and $\xi(t)\in \lbrace -1,0,1\rbrace$. In the case that the base linear system is stable and the entire flow region belongs to the converging region, $\beta = 1$ can be selected. Else $\beta = 0$ is to be used in which case, $[\underline{\tau}, \overline{\tau}]$ is defined over part of the flow region.
\begin{equation}
	\xi = \sign{\dfrac{s}{s/\omega_e + 1}\norm{e}}
\end{equation}
For large values of the corner frequency $\omega_e$, update is enabled for increasing error. Where $\omega_c$ is the bandwidth of the closed loop, values for $\omega_e>>\omega_c$ result in estimating the lower bound on the converging region to be $\underline{\tau}=0$, which is generally not true. Also large values for $\omega_e$ will increase the influence of noise. Decreasing the value of $\omega_e$ results in increased phase lag, which creates a good estimate of the region $[\underline{\tau}, \overline{\tau}]$. A properly chosen value for $\omega_e$ ensures stable adaptation. As a rule of thumb, values for $\omega_e\in[\frac{\omega_c}{10}, \omega_c]$ are desirable.

\section{Application on Precision Positioning Stage} \label{sec:application}

The adaptive algorithm is tested on a precision positioning stage. Performance of the algorithm can be tested and expressed in the convergence of the feedforward parameters, which should also result in a corresponding reduction in tracking error. In an ideal environment, the feedforward controller provides all the control action taking over from the feedback element. However, in practice plants are constrained to model mismatches and external disturbances, which requires a feedback element to provide part of the control action. 

\subsection{Setup}
\begin{figure}[htbp]
	\centering
	\includegraphics[width=0.35\textwidth]{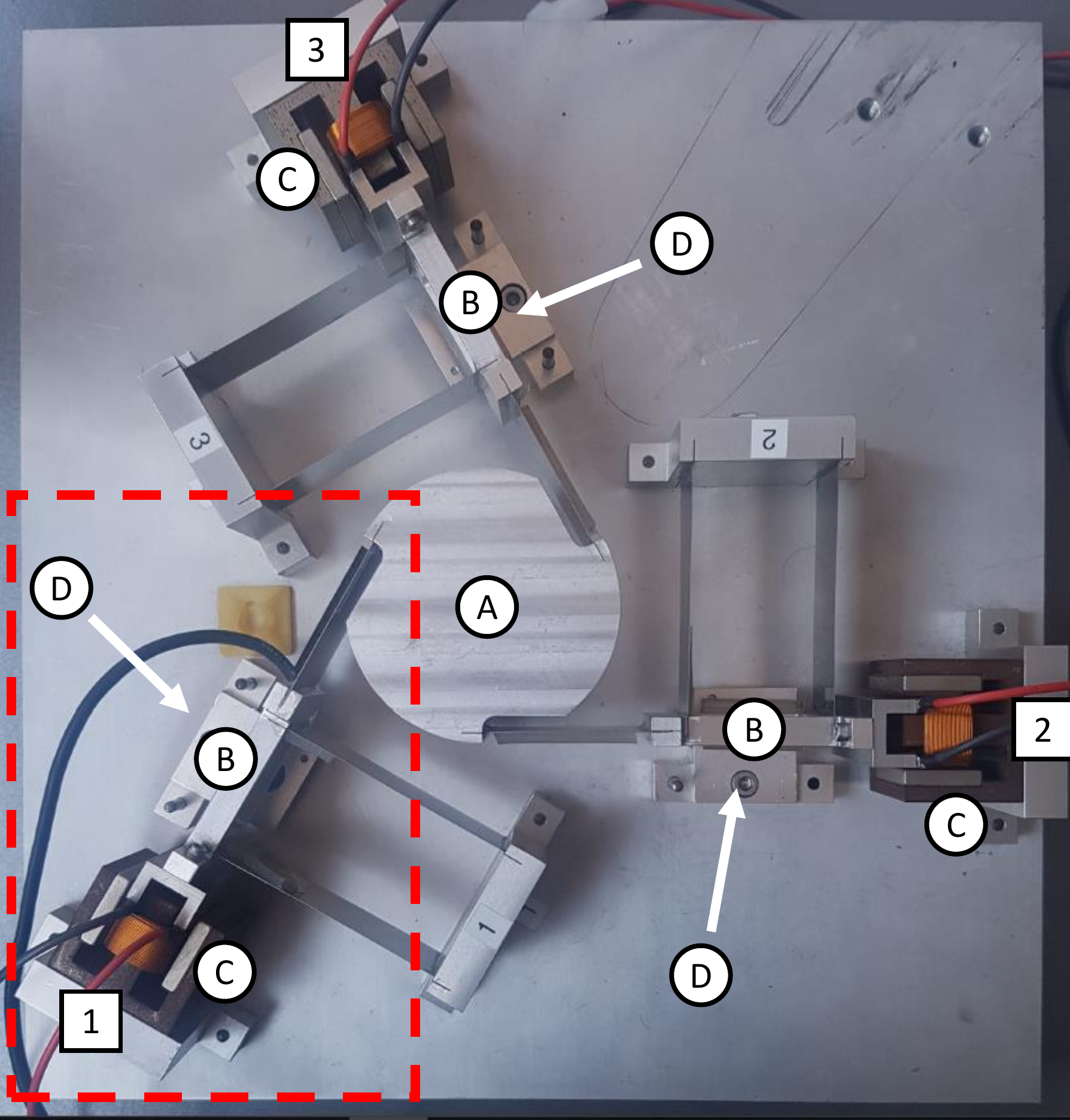}
	\caption{Spider positioning stage}
	\label{fig:spyder}
\end{figure}

The control structure is validated on the planar precision motion stage named the Spider stage, shown in \fref{fig:spyder}. Three masses (B) are actuated by individual voice coils (C). The masses are constrained to the base plate by leaf flexures. The position of the individual masses is measured by a linear encoder (D). By controlling the three individual masses, the centre mass (A), which is connected to the three individual masses by a leaf flexure,  can be accurately positioned. For performance measurement of the feedforward algorithm, the stage is simplified as a SISO plant by only actuating and measuring one of the three masses. Only mass 1, indicated by the red square in \fref{fig:spyder}, is controlled. The controller is implemented on the FPGA of an NI CompactRIO with a sampling frequency of \SI{10}{ kHz}. The encoder provides a resolution of \SI{100}{nm}.

\begin{figure}[htbp]
	\centering
	\includegraphics[trim = {0.5cm 0cm 1.5cm 0cm}, width=\linewidth]{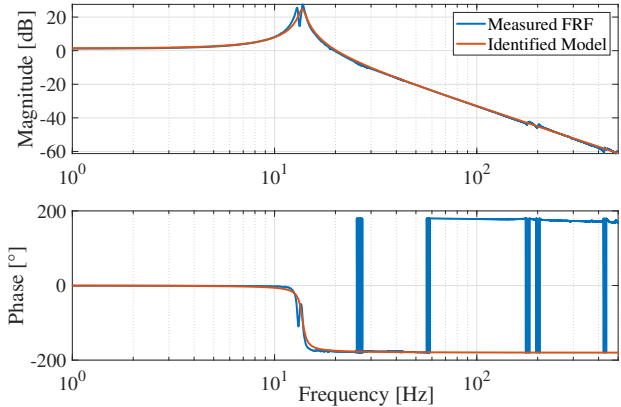}
	\caption{Measured frequency response of the Spider stage fitted with the frequency response of the second-order estimated plant model for 1-500 Hz}
	\label{fig:Identification}
\end{figure}

Identifying the system, its dynamics can be represented by a double mass-spring-damper system with the two modes close to each-other. However, to prove the effectiveness of the adaptive algorithm, an approximation as a single mass-spring-damper system suffices. The transfer function of the second-order system is given below.
\begin{equation}
	P(s) = \dfrac{8760}{s^2 + 5.886s + 7443}
	\label{eq:P2}
\end{equation}
The obtained frequency response is fitted with the frequency response of the second-order plant of \eref{eq:P2} as shown in \fref{fig:Identification}. Even though this is a reasonable representation of the dynamics, the plant shows slight non-linear behaviour for the stiffness and damping terms.

\subsection{Feedback Controller Configurations}
All feedback controllers are designed for a bandwidth of $\omega_c = 100$ Hz. Since the convergence of the feedforward parameters for different types of controllers is to be validated, the feedback controllers are not designed for performance comparison. Four different controllers are tested; linear PID, CI based PID, and two different CgLp based PID controllers. The phase margins for the different controllers are obtained using a describing function analysis.
\subsubsection{PID}
As the update algorithm is designed for reset controllers, the effectiveness is also noticeable for application of linear PID controllers. For precision industry, a general series PID for loop-shaping is given by \eref{eq:PID}. This linear controller is also the basis for the other reset controllers.
\begin{equation}
	C(s) = k_p\left(\dfrac{s+\omega_i}{s}\right)\left(\dfrac{1+\frac{s}{\omega_d}}{1+\frac{s}{\omega_t}}\right)\left(\dfrac{1}{1+\frac{s}{\omega_f}}\right)
	\label{eq:PID}
\end{equation}
with $\omega_i<\omega_d<\omega_t<\omega_f$. The linear PID controller is designed for a phase margin of approximately 41\textdegree. The different frequencies are obtained from \cite{mechatronics} using the rule of thumb as $\omega_i = \omega_c/10$, $\omega_d = \omega_c/3$, $\omega_t = 3\omega_c$ and $\omega_f = 10\omega_c$.
\subsubsection{CI based PID}
CI shows its advantage in reduction of phase lag and hence the linear integrator is replaced with the Clegg Integrator. In reset control, the CI based controller shows the most abrupt changes in control value which impairs convergence of the feedforward parameters.   
\begin{equation}
	C(s) = k_p\left(\dfrac{s+\omega_i}{\cancelto{}{s}}\right)\left(\dfrac{1+\frac{s}{\omega_d}}{1+\frac{s}{\omega_t}}\right)\left(\dfrac{1}{1+\frac{s}{\omega_f}}\right)
\end{equation}
The controller provides a phase margin of approximately 51\textdegree \ according to describing function analysis with $\omega_i = \omega_c/10$, $\omega_d = \omega_c/1.2$, $\omega_t = 1.2\omega_c$ and $\omega_f = 10\omega_c$.
\subsubsection{CgLp based PID 1}
The CgLp element can be added to the controller to provide broadband phase compensation. The controller is designed as below for a phase margin of approximately 47\textdegree. The CgLp element, however, shows less abrupt changes in control value at reset instants compared to CI based PID. 
\begin{equation}
	C(s) = k_p\left(\dfrac{\frac{s}{\omega_r}+1}{\cancelto{}{\frac{s}{\omega_{r\alpha}}+1}}\right)\left(\dfrac{s+\omega_i}{s}\right)\left(\dfrac{1+\frac{s}{\omega_d}}{1+\frac{s}{\omega_t}}\right)\left(\dfrac{1}{1+\frac{s}{\omega_f}}\right)
\end{equation}
\begin{equation*}
	\begin{aligned}
		\omega_i &= \omega_c/10 & \omega_d &= \omega_c/1.4 & \omega_t &= 1.4\omega_c \\
		\omega_f &= 10\omega_c & \omega_{r} &= \omega_c/4 & \omega_{r\alpha} &= \omega_r/1.4 
	\end{aligned}
\end{equation*}
\subsubsection{CgLp based PID 2}
The amount of non-linearity is determined by either providing phase through the reset element or the lead filter. Comparison of convergence for two different CgLp configurations shows the effectiveness of the adaptive algorithm to different types of reset controllers. A second CgLp element with low non-linearity is validated in experiments. As the necessary phase is dominantly provided through the linear element, the converging region of the base linear system is larger. The overall phase margin is 52\textdegree.
\begin{equation*}
	\begin{aligned}
		\omega_i &= \omega_c/10 & \omega_d &= \omega_c/2.5 & \omega_t &= 2.5\omega_c \\
		\omega_f &= 10\omega_c & \omega_{r} &= \omega_c/1.1 & \omega_{r\alpha} &= \omega_r/1.4 
	\end{aligned}
\end{equation*}

\subsection{Simulations}
\begin{figure}[tb]
	\centering
	\includegraphics[trim = {1.5cm 0cm 1.5cm 0cm}, width=\linewidth]{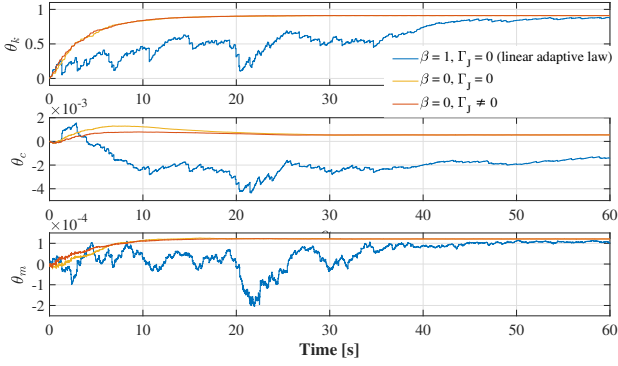}
	\caption{Evolution of $\theta_k$, $\theta_c$, and $\theta_m$ using a CI based reset feedback controller in a simulation, showing the stabilizing effect of the novel adaptive algorithm compared to a linear feedback learning controller, and the faster convergence using the jump update}
	\label{fig:comparelinear}
\end{figure}

The effectiveness of the proposed adaptive algorithm compared to linear methods can be easily demonstrated by setting $\beta=1$ in \eref{eq:psi}. By also disabling the update of the feedforward parameters in the jump set, the adaptive algorithm is linear. To show both the need for defining the converging region and the use of introducing a jump update, three different simulations in MATLAB are done. In this comparison the following three settings are used in a simulation of 60 seconds:
\begin{enumerate}
	\item $\beta = 1$, $\Gamma_J = 0$  (Pure linear adaptive law)
	\item $\beta = 0$, $\Gamma_J = 0$  (Proposed law without jump update)
	\item $\beta = 0$, $\Gamma_J \neq 0$  (Proposed adaptive law)
\end{enumerate}  
In the simulations, a noise and disturbance-free case is considered. The reference consists of three added sinusoids, the same as used in the physical experiments. The adaptive gain matrices for the comparison in the simulation are listed in \Cref{tab:gamma} under CI. The results of the simulation are shown in \fref{fig:comparelinear}. It becomes clear that the converging region needs to be defined for stable adaptation. Also, by enabling the jump update, faster adaptation is achieved in the starting region.

\subsection{Experiments}
In the experiments, the system is sampled at a rate of \SI{10}{kHz}. All linear filters are discretized for the sampling rate using the bilinear Tustin approximation, this includes the feedback controllers and error filter $\psi$. Euler's approximation is used for updating the feedforward parameters.
\begin{equation*}
	\theta(k+1) = \theta(k) + T_s\dot{\theta}(k)
\end{equation*}
The jump update is discrete by definition, therefore this can be implemented as presented in \eref{eq:updatelaw}. Since the reference is known beforehand, direct derivatives are used for $\phi$. The system can be modelled as a second-order plant, therefore a feedforward of only order two is used. This yields that feedforward parameters relatively represent the stiffness, damping, and mass.
\begin{equation*}
	\theta = \begin{bmatrix}
		\theta_k & \theta_c & \theta_m
	\end{bmatrix}^T \qquad \phi = \begin{bmatrix}
		r & \dot{r} & \ddot{r}
	\end{bmatrix}^T
\end{equation*}
Since the feedforward consists of three parameters, a multi-sine signal with the sum of at least two sinusoids suffices. Less periodicity is created by adding another sine of a frequency which is indivisible by the other two frequencies. 
\begin{equation*}
	r(t) = A\left(sin\left(2\pi t\right)+sin\left(10\pi t\right)+sin\left(28\pi t\right)\right)
\end{equation*}
The adaptive feedforward is tested for the four different feedback controller cases. Additional experiments have been done on the CgLp based PID 1 controller with artificially added sinusoidal disturbance in \SI{}{mV}.
\begin{equation*}
	d(t) = 15.3\sin\left(30\pi t\right)
\end{equation*}

\subsection{Results}
$\beta=0$ is used for all controllers. Any initial values for $\theta$ should theoretically show convergence. Since it is only desired to prove convergence and hence reduction in tracking error, $\theta_0$ is chosen as a zero vector. Since the same reference is applied in every case, it is straightforward to use the same normalization matrix $P = \diag{1000, 1, 1}$. As $\xi(t)$ determines the converging region, $\omega_e$ has to be set correctly. For all feedback controllers, approximately the same phase margin and the same bandwidth is used, therefore $\omega_e$ is kept constant through all experiments at $\omega_e = \omega_c/10$.

\begin{figure}[tb]
	\centering
	\includegraphics[trim = {1.5cm 0cm 1.5cm 0cm}, width=\linewidth]{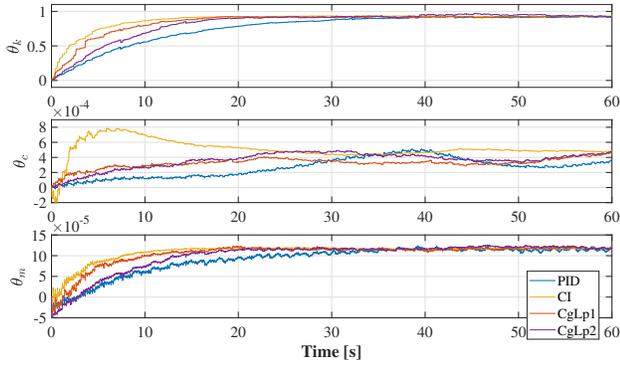}
	\caption{Evolution of $\theta_k$, $\theta_c$, and $\theta_m$ for the different feedback controllers}
	\label{fig:All}
\end{figure}

The adaptive gain matrices for the different controllers are listed in \Cref{tab:gamma}. Note that the gains are more or less correlated with the relative difference of the second-order plant parameters. Experiments are run for 60 seconds to properly show convergence and steady-state behaviour of the feedforward parameters. The results for the four different controllers are shown in \fref{fig:All}.
\begin{table}[]
	\centering
	\caption{Values of $\Gamma_F$ and $\Gamma_J$ for the different feedback controllers}
	\begin{tabular}{|l|l|l|}
		\hline
		controller & $\Gamma_F$ & $\Gamma_J$ \\ \hline
		PID & $\diag{150000, 5, 1}$ & $\diag{0, 0, 0}$ \\ \hline
		CI & $\diag{40000, 2, 0.2}$ & $\diag{2, 0.0002, 0.00002}$ \\ \hline
		CgLp 1 & $\diag{100000, 3, 0.5}$ & $\diag{5, 0.0001, 0.00005	}$ \\ \hline
		CgLp 2 & $\diag{150000, 4, 0.5}$ & $\diag{5, 0.00005, 0.00005}$ \\ \hline
	\end{tabular}
	\label{tab:gamma}
\end{table} 

The time response $\theta_k$, $\theta_c$, and $\theta_m$ of all controller configurations are presented in one figure. As expected, in all cases $\theta_m$ converges to the same value as this represents the mass of the plant, which was constant during experiments. The damping value $\theta_c$ varies with respect to position and velocity and hence varies over time, depending on the magnitude of the same parameters in reference over time. As damping is a non-linear phenomenon modelled as linearity in the system, this behaviour is expected. The stiffness, represented by $\theta_k$ stays more constant but changes at the same rate as $\theta_c$. The non-linearity of the stiffness is not observed very clearly as the deflection stays within a linear region due to the bounded reference. 

\begin{figure}[htb!]
	\centering
	\includegraphics[trim = {1.5cm 0cm 1.5cm 0cm}, width=\linewidth]{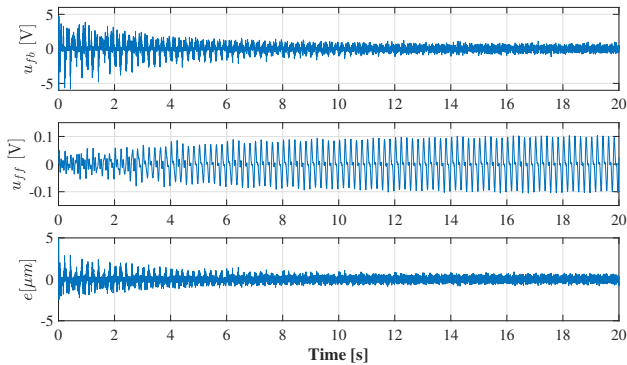}
	\caption{Evolution of $u_{fb}$, $u_{ff}$ and $e$ for the first 20 seconds of the CgLp based PID controller 1}
	\label{fig:signals}
\end{figure}

\begin{figure}[htb!]
	\centering
	\includegraphics[trim = {1.5cm 0cm 1.5cm 0cm}, width=\linewidth]{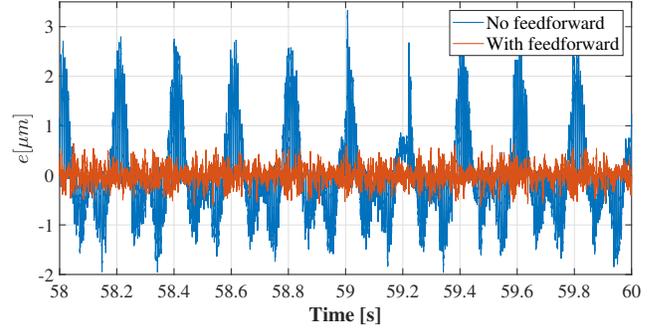}
	\caption{Comparison of the reference tracking error $e$ with and without the converged feedforward controller at steady state, using the CgLp based PID controller 1}
	\label{fig:error1}
\end{figure}

The goal of the feedforward is taking over control action of the feedback controller to decrease the reference tracking error. For the CgLp base PID controller 1, the feedback control input $u_{fb}$, feedforward control input $u_{ff}$, and tracking error $e$, are shown in \fref{fig:signals}. These three signals are similar for all feedback controllers since all show convergence of the feedforward parameters. The signals are only shown for the first 20 seconds as the feedforward parameters are converged within this time interval. It is observed that the feedback controller plays a significant role at the start of the experiment and decreases in value as the time increases. The value does not converge to zero; this can mainly be explained by the presence of noise and aforementioned plant non-linearities. The reference tracking error is decreased by approximately 75\% over time in this specific case. Comparison of tracking error for a control structure without feedforward and one with adapted feedforward is shown in \fref{fig:error1}. In this case, a reduction in error of approximately 85\% is observed, making the feedforward necessary for accurate tracking.


\begin{figure}[htb!]
	\centering
	\includegraphics[trim = {1.5cm 0cm 1.5cm 0cm}, width=\linewidth]{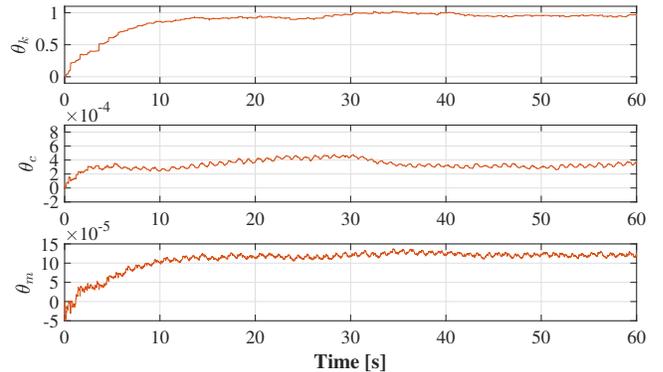}
	\caption{Evolution of $\theta_k$, $\theta_c$, and $\theta_m$ of the CgLp based PID controller 1 with applied disturbance with an amplitude of 50 at 15 Hz}
	\label{fig:CgLpDist}
\end{figure}

\begin{figure}[htb!]
	\centering
	\includegraphics[trim = {1.5cm 0cm 1.5cm 0cm}, width=\linewidth]{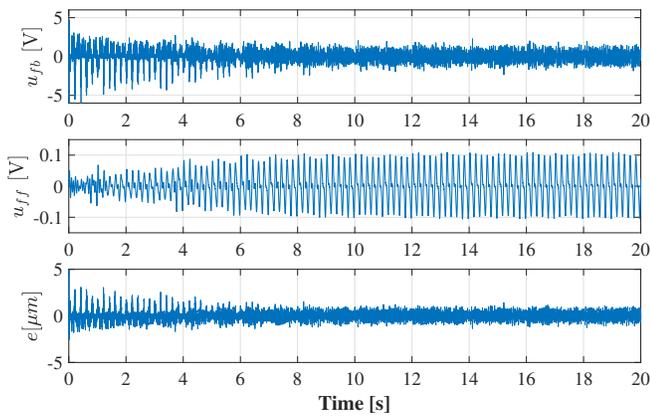}
	\caption{Evolution of $u_{fb}$, $u_{ff}$ and $e$ for the first 20 seconds of the CgLp based PID controller 1 with applied disturbance with an amplitude of 15.2 mV at 15 Hz}
	\label{fig:signalsDist}
\end{figure}

\begin{figure}[htb!]
	\centering
	\includegraphics[trim = {1.5cm 0cm 1.5cm 0cm}, width=\linewidth]{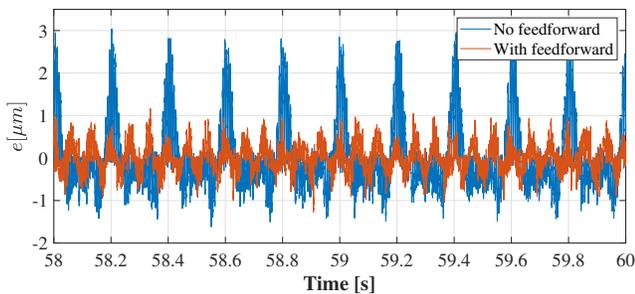}
	\caption{Comparison of the reference tracking error $e$ with and without the converged feedforward controller at steady state, using the CgLp based PID controller 1. With applied disturbance with an amplitude of 15.2 mV at 15 Hz}
	\label{fig:error2}
\end{figure}

To test the algorithm for robustness to disturbance, a final test is done with added disturbance. A relatively high amplitude sinusoidal input disturbance is artificially added to the control signal. The same controller configuration of CgLp based PID 1 is used. The time response of $\theta$ is shown in \fref{fig:CgLpDist}. It is observed that feedforward parameters converge as in the disturbance-free case. However, all $\theta$ oscillate at steady state. The disturbance is applied at a frequency which is nearly at resonance. To reject this disturbance, the feedback controller has to counteract with a high amplitude control signal, which inherently causes the feedforward parameters to slightly oscillate. However, the feedforward parameters oscillate around the ideal value and do not diverge. \fref{fig:signalsDist} shows a decrease of the feedback control input and error as the feedforward control input increases. The tracking error for this case decreases by approximately 65\% over time. \fref{fig:error2} shows the difference in tracking error for a case with and without feedforward. An error reduction of 65\% is observed here. Note that error without additional feedforward is approximately the same as in the disturbance-free case of \fref{fig:error1}. This shows the effectiveness in disturbance rejection of the reset controller. 


\section{Conclusion} \label{sec:conclusion}

Reset controllers provide a good alternative for linear PID control because of an increase in stability margins and the ability to use loop shaping techniques. The reference tracking properties of a reset controller are not always good. Parallel feedforward control can be used to improve for accurate reference tracking. The accuracy of a fixed feedforward is directly dependent on the correctness of the identified plant. For increased precision, adaptive or iterative models can be used which estimate the parameters of the feedforward controller on-line. Existing linear update algorithms are not suited for reset control due to the non-linear nature of the control signal. In this research, an adaptive feedforward algorithm is proposed which can be applied with any type of reset feedback controller based on zero crossings of the tracking error. The scheme is based on linear parametrized plant inverse. The adaptive algorithm consists of both a flow update and a jump update for increasing the rate of convergence. Experimental results have shown convergence of the feedforward parameters for a mass-spring-damper system, with a PID, CI based, and CgLp based controller in the feedback loop, as well with added disturbance.

\bibliographystyle{IEEEtran}
\bibliography{bibliography}

\end{document}